\documentclass[12pt,a4paper]{article}
\usepackage{graphicx}
\usepackage{epsfig}
\def\b{\bar}

\def\m{\mu}
\def\n{\nu}

\def\~{\widetilde}

\def\bY3{\bar Y_{,3}}
\def\Y3{Y_{,3}}
\def\z{\zeta}
\def\Z{{\b\zeta}}
\def\Y{{\bar Y}}

\def\`{\dot}
\def\be{\begin{equation}}
\def\ee{\end{equation}}
\def\bea{\begin{eqnarray}}
\def\eea{\end{eqnarray}}

\def\fn{\footnote}

\def\mn{{\mu\nu}}

\begin{document}
\textwidth=135mm
 \textheight=200mm
\begin{center}
{\bfseries What tells Gravity on the  shape and size of an
 electron \footnote{{\small Talk at the SPIN2012 Conference,
JINR, Dubna, September 17 - 22, 2012.}}} \vskip 5mm A. Burinskii
\vskip 5mm {\small {\it Theor.Phys.Lab., NSI Rus. Acad. Sci.,
B.Tulskaya 52, 115191 Moscow, Russia}}
\\
\end{center}
\vskip 5mm \centerline{\bf Abstract} Gravitational field of an
electron, fixed by experimental values of its mass, spin, charge
and magnetic moment, is given by the metric of Kerr-Newman (KN)
solution. Unexpectedly, this metric contains a singular ring of
the Compton radius, which should be regulated resulting in a weeak
and smooth source.
The consistent source takes the form of an oblate vacuum bubble,
bounded by a closed string of the Compton radius. The bubble
turns out to be relativistically rotating and should be filled by a coherently
oscillating Higgs field in a false vacuum state.

 \vskip 10mm
{\bf Introduction.} It is commonly recognized now that black holes are
akin to elementary particles. The Kerr-Newman solution has
gyromagnetic ratio $g=2$ as that of the Dirac electron, and the
experimentally observable parameters of electron determine its
asymptotical gravitational field in accord with the Kerr-Newman solution.
The spin/mass ration of the electron is extremely high,  $J/m
\sim 10^{22} $ (we use the units $G=c=\hbar =1$), and the
\emph{black hole horizons disappear}, opening the naked Kerr singular ring
 of the Compton radius $\sim 10^{-11}$ cm. It is very far from the expected
 pointlike electron of quantum theory. Besides, quantum theory
 supposes a flat minkowskian background, and this singular region
 should be regularized by some procedure leading to a finite and
 smooth source of the KN solution with a flat metric in vicinity of the
 electron core.  It is not
 a priory clear that such a source can be obtained, and the aim of this paper is
 to describe basic elements of the given in \cite{BurSol} electron model
 which is  consistent with the external KN solution and the above mentioned
 quantum requirements.

{\bf Structure of the KN solution.} Metric of the KN solution has
the form
 \be g_\mn=\eta _\mn + 2 H k_\m k_\n , \qquad H=\frac {mr -
e^2/2}{r^2 + a^2 \cos ^2 \theta}, \label{ksH}\ee
 where  $\eta _\mn $
is metric of auxiliary Minkowski space in the Cartesian
coordinates $\rm x\equiv(t, x, y, z) \in M^4,$ and $k^\m(\rm x)
\in M^4$ is a lightlike vector field, forming a twisting congruence
shown in  Fig.1. Coordinates $r, \theta$ and $\phi_K$ are Kerr's oblate spheroidal
coordinates (Fig.2). The KN metric is singular at the circle $r
=\cos \theta =0 ,$ which is \emph{branch line of the Kerr space
into two sheets $r^+$ for $r>0$ and $r^-$ for $r<0 ,$}  so that the field $k^{\m}(\rm
x)$ and the aligned with $k^\m $ metric and vector potential of the
electromagnetic (em) field, \be \alpha^\m_{KN} = Re \frac e {r+ia
\cos \theta} k^\m \label{ksGA} ,\ee turn out to be twosheeted, taking different
values on the different sheets of the same point $\rm x \in M^4 .$
Twosheetedness represents one of the main puzzles of the KN space-time.
 For electron parameters, gravitational field of the KN solution
is concentrated very close to singular ring, forming a circular
waveguide -- analog of the closed relativistic
string. It has been shown in \cite{BurSen,BurQ} that the KN
solution in vicinity of the Kerr ring corresponds to the obtained
by Sen solution to low-energy heterotic string theory. Meanwhile,
the long-term attack on the mysterious twosheetedness (Keres,
Israel, Hamity, L\'opez at all, \cite{BurQ}) resulted in the
gravitating soliton model in the form of the consistent with KN
solution rotating vacuum bubble, metric of which is
regularized, approaching the flat minkowskian background in the
Compton region.  It fixes unambiguously the form and some details of the consistent
with KN gravity electron model.
\begin{figure}[h]
\begin{minipage}{17pc}
\includegraphics[width=17pc]{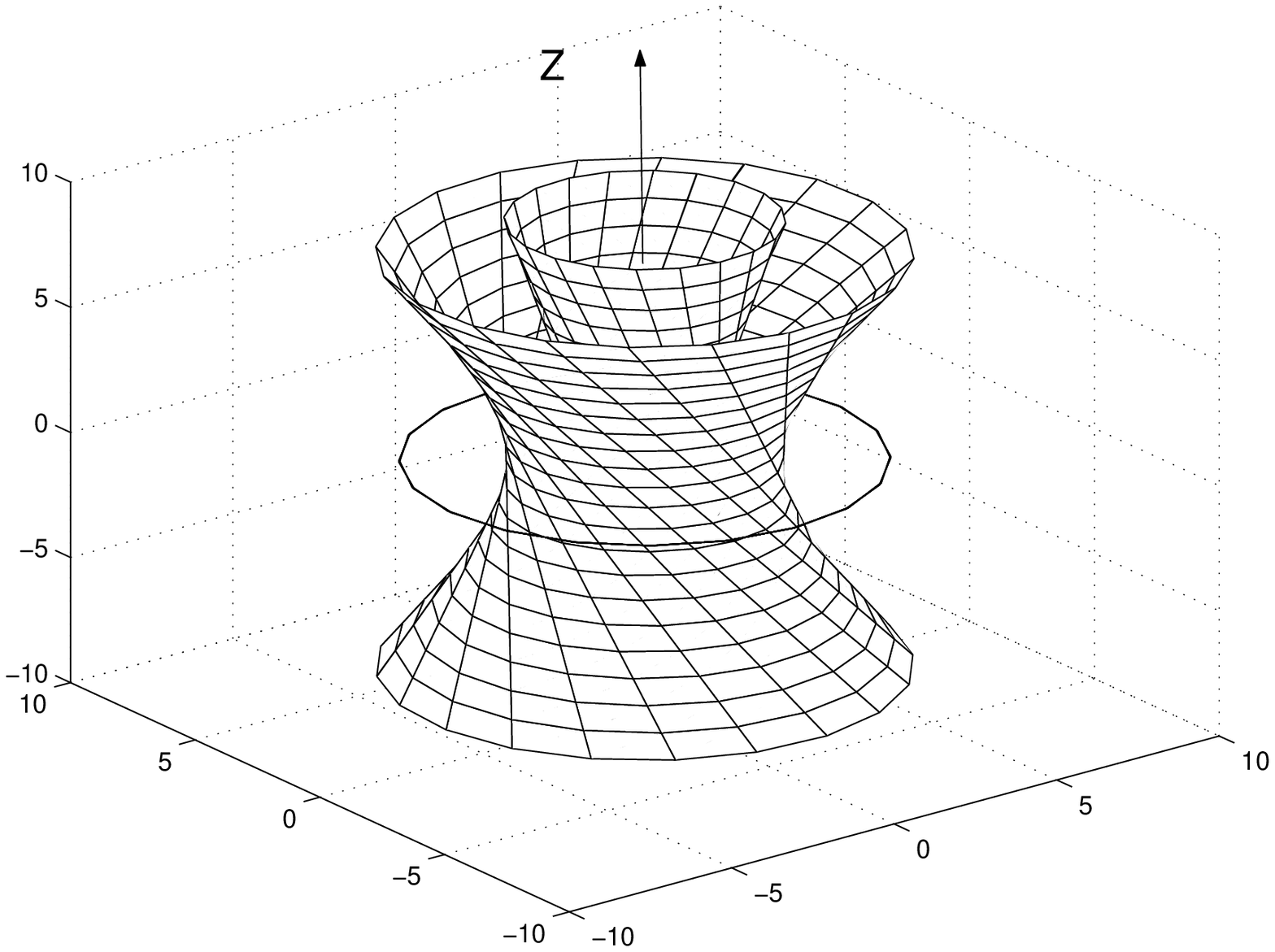}
\caption{\label{label}Congruence of the lightlike lines $k^\m(x)$
is focused on singular ring, creating twosheeted Kerr space.}
\end{minipage}\hspace{4pc}%
\begin{minipage}{10pc}
\includegraphics[width=10pc]{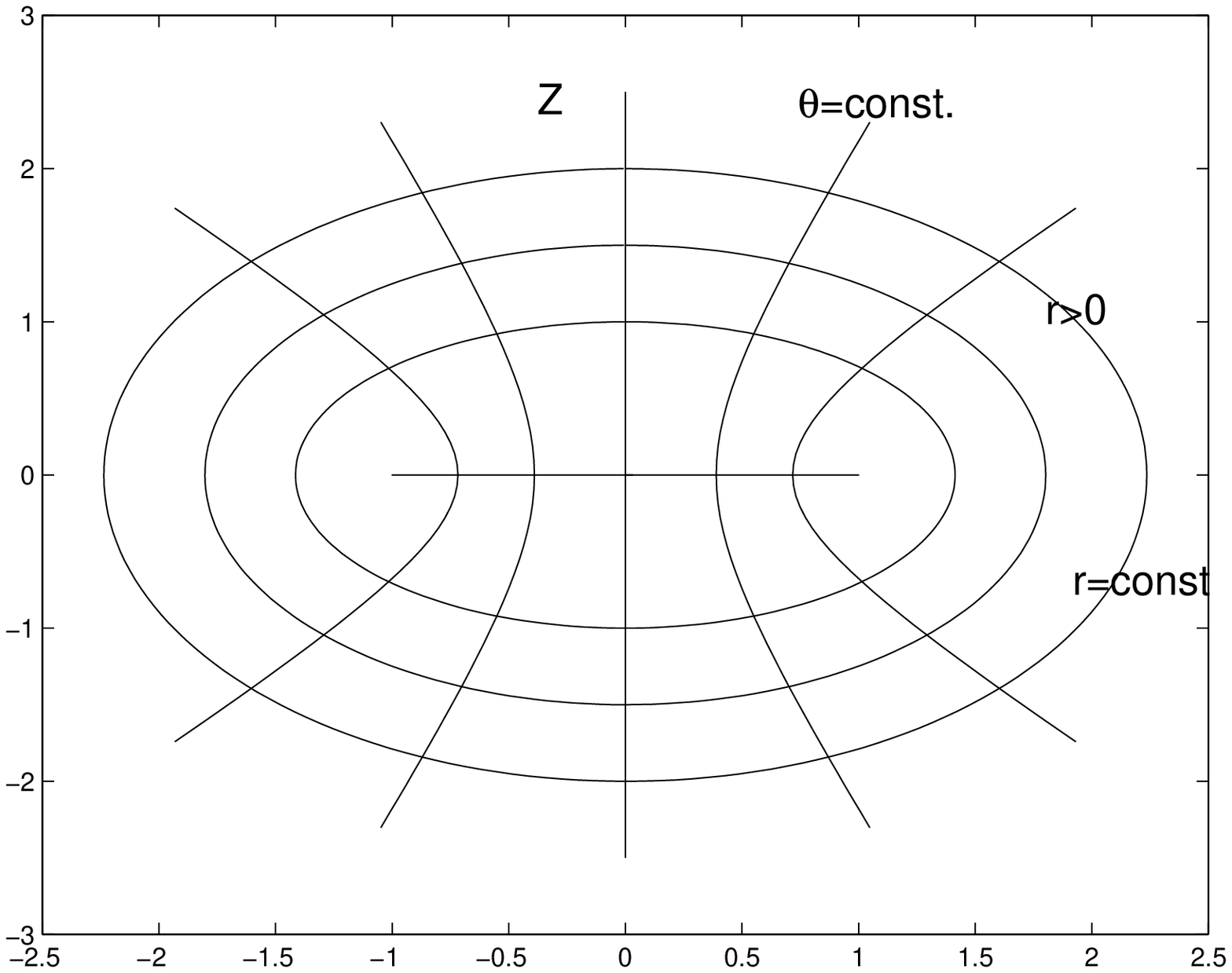}
\caption{Oblate coordinate system $(r,\theta)$ covers the Kerr
space twice, for $r>0$ and $r<0.$ Truncation of the sheet $r<0$
creates the source at $r=0.$ }
\end{minipage}
\end{figure}
Following \cite{BurSol}  we discuss basic features of the
regular KN electron model as a gravitating soliton. The most
wonderful fact is emergence of the quantum condition for spin of
the KN soliton, as a consequence of the pure classical relations
completed by the condition on periodicity of the Wilson loop
(analog of the Bohm-Aharonov effect). We obtain also that excitations
of circular string  deform  boundary of the bubble and create a
circulating  singular pole, exhibiting as
``zitterbewegung'' of the Dirac electron.

{\bf Solitonic source of the regularized KN geometry.}
Differential form $k_\m dx^\m  = P^{-1}( du +
\bar Y d \zeta + Y d \bar\zeta - Y \bar Y dv)$ determines the null
vector field $k^\m$ in the \emph{null Cartesian} coordinates $ \z =
(x+iy)/\sqrt 2 ,\quad  \Z = (x-iy)/\sqrt 2 , u = (z + t)/\sqrt 2
,\quad v = (z - t)/\sqrt 2 .$ Function
$Y(\rm x)$ is determined by the Kerr theorem, and in the Kerr angular
coordinates it takes the form $Y(\rm x) = e^{i\phi_K} \tan \theta/2 .$

Using relation between the
Kerr coordinates  $r, \theta, \phi_K$ and Cartesian ones
\be x+iy = (r + ia) \exp \{i\phi_K \} \sin \theta , \quad
z=r\cos \theta, \quad \rho =r-t , \label{coordKerr} \ee
one obtains the form
$k_\m dx^\m=dt +\frac z r dz + \frac {r (xdx+ydy)}{r^2 +a^2} -
\frac {a (xdy-ydx)}{r^2 +a^2} ,$
which shows that in the
equatorial plane ($z=\cos \theta=0$), the Kerr congruence forms a vortex,
approaching the Kerr ring tangentially \be k
|_{r=\cos \theta=0}= dt - (xdy-ydx)/a = dt -a d\phi . \label{k
tangent}\ee Therefore, the Kerr ring is lightlike, similar to the
DLCQ circle of M-theory.
Truncation of the `negative' Kerr sheet (H.Keres, 1967; W.Israel,
1968)  creates a disklike source, $r=0,$ spanned by the Kerr ring.
V.Hamity (1976) showed  that the disklike source represents a
rigid and relativistically rotating membrane, with the lightlike
boundary.

\textbf{The regular bubble model} was proposed by L\'opez, which
suggested to truncate the negative sheet along the ellipsoidal
shell $r=r_e>0 ,$ which covers the singular ring. There appears a
rotating bubble source with a flat minkowskian interior.
 One sees that the external KN metric \ref{ksH} is matched with
 the flat interior by the condition $H=0,$ which fixes the
 boundary of the bubble source at \be r= r_e= e^2/(2m) \label{rb} .\ee
 It follows from (\ref{coordKerr}) that \emph{the bubble takes the form of a
highly oblate ellipsoid of the Compton radius $a \approx \hbar/m $
and thickness  $ 2r_e = e^2/m .$}\fn{The exact value of the disk
radius is $r_b = \sqrt{a^2 + r_e^2} \approx a + \frac 12 \delta ,$
where $ \frac \delta a \approx 3\cdot10^{-46}.$}

{\bf Gravitating soliton.} In \cite{BurSol}, the bubble shell
model was extended to a smooth field
 model of the domain wall bubble interpolating between the
external KN background and a false vacuum state inside the bubble
(for details see \cite{BurSol}).  The Kerr singular ring is
suppressed by  \emph{the supersymmetric vacuum state with a flat
metric inside the bubble,} the Kerr closed string is formed by
the em field concentrating at the edge rim of the
bubble.

\textbf{Regularization of the em field.} Regularization is
performed by the Higgs mechanism of broken symmetry. One of the
complex chiral fields, say $\Phi$ is considered as a Higgs field,
which takes the non-zero vev $\Phi = \Phi_0 \exp(i\chi )$ inside
the bubble and pushes out the em field. The typical Lagrangian
yields the equation \be \chi,_\m + e \alpha_\m =0 \label{Main}
.\ee
 The cut-off $r_e$ determines  maximum of the KN vector potential

 $\alpha^{str} = \alpha^{max}_{KN} = e/ r_e = 2m/e ,$ which
is concentrated in the
 form of a closed string at the edge border of the disklike bubble. In
agreement with (\ref{k tangent}),  the longitudinal component of
$\alpha^{str}$ forms a closed Wilson loop along the boundary of
the disk, which yields $ \oint e\alpha_\m^{str} d l^\m = 4\pi
ma .$ Using Kerr's relation $J=ma ,$ we obtain for the loop integral of
(\ref{Main}) over the disk boundary
\be \oint \chi,_\m dl^\m = 2\pi n = - 4 \pi J , \label{oint}\ee
which indicates \emph{quantization of the spin-projection,} $|J|= n/2 , \quad n=1,2...$

Similarly, the time component of (\ref{Main}) yields $\dot \chi= \omega
 =2m ,$  resulting in \emph{oscillations of the Higgs
 field with the frequency $2m ,$} what is
 typical for the "oscillon" type of the solitonic  models.

It was also revealed recently, that em traveling waves, propagating along
the Kerr circular string, should deform the bubble boundary and break regularity
condition at some node which moves together with string excitations. It results in
emergence of a circulating singular pole, which exhibits ``zitterbewegung'' of
the Dirac electron. Therefore, along with the prompted by KN gravity disk-like shape of
the``dressed''  electron and circular string of Compton radius,
excitations of this string create point-like structure of the``naked'' electron,
which appears in the form of the circulating singular pole.

\end{document}